\documentclass[preprint,pra,superscriptaddress]{revtex4-1}

\usepackage{physics,graphicx,amsmath,amssymb,amsthm,subfig,xcolor,tikz}

\usepackage[ruled]{algorithm2e} \usepackage{algpseudocode}

\usepackage[colorlinks, linkcolor=red, anchorcolor=blue,
citecolor=green]{hyperref} \usepackage{color}
\DeclareMathOperator*{\argmin}{arg\,min}

\begin{document}

\title{Group theory on quantum Boltzmann machine}

\author{Hai-Jing Song}

\affiliation{Institute of Physics, Beijing National Laboratory for Condensed
  Matter Physics, Chinese Academy of Sciences, Beijing 100190, China}

\affiliation{School of Physical Sciences, University of Chinese Academy of
  Sciences, Beijing 100049, China}

\author{D.~L. Zhou}

\email{zhoudl72@iphy.ac.cn}

\affiliation{Institute of Physics, Beijing National Laboratory for Condensed
  Matter Physics, Chinese Academy of Sciences, Beijing 100190, China}

\affiliation{School of Physical Sciences, University of Chinese Academy of
  Sciences, Beijing 100049, China}

\affiliation{Collaborative Innovation Center of Quantum Matter, Beijing 100190,
  China}

\affiliation{Songshan Lake Materials Laboratory, Dongguan, Guangdong 523808,
  China}

\begin{abstract}
  Group theory is extremely successful in characterizing the symmetries in
  quantum systems, which greatly simplifies and unifies our treatments of
  quantum systems. Here we introduce the concept of the symmetry for a quantum
  Boltzmann machine and develop a group theory to describe the symmetry. This
  symmetry implies not only that all the target states related with the symmetry
  transformations are equivalent, but also that for a given target state all the
  optimal solutions related with the symmetry transformations that keeps the
  target state invariant are equivalent. For the Boltzmann machines built on
  qubits, we propose a systematic procedure to construct the group, and develop
  a numerical algorithm to verify the completeness of our construction.
\end{abstract}

% \pacs{xxx}

\maketitle

\section{Introduction~\label{sec:intro}}

Quantum Boltzmann machine is one of the major machine learning models in quantum
machine learning~\cite{RN1, RN2, RN9}. A quantum Boltzmann machine is a quantum
system with a specific type of Hamiltonians with variable parameters, and it is
controlled to approximate a target quantum state as well as possible via tuning
these parameters~\cite{RN3, RN4, RN6, Nihal08}. The target quantum state is a
``quantum data'', which may contain quantum characteristics such as quantum
coherence or entanglement that can not be simulated classically. In
particularly, a quantum Boltzmann machine is proposed to be implemented in a
D-Wave anneal device, which makes it possible to be trained by experimental
samplings~\cite{RN5, RN6}.

Symmetry plays a predominant role in quantum physics, where it is successively
described by group theory~\cite{Ste1994, RN7, RN10}. However, group theory has
rarely been introduced to study quantum Boltzmann machine. Recently we have
found one example where symmetry can be used in a specific quantum Boltzmann
machine~\cite{RN4}. Here we aim to extend our case study~\cite{RN4} to a group
theory suitable for any quantum Boltzmann machine. First we will clarify the
concept of the symmetry for a quantum Boltzmann machine, and introduce the
symmetry group of a quantum Boltzmann machine to describe the symmetry. Then we
answer what the symmetry implies for a quantum Boltzmann machine.

Further more, we focuses on how to construct the symmetry group for a general
quantum Boltzmann machine. We obtain a systematic procedure to construct the
symmetry group of a quantum Boltzmann machine, and develop a numerical algorithm
to show the completeness of our construction.

\section{Definition of quantum Boltzmann machine~\label{sec:defin-quant-boltzm}}

A quantum Boltzmann machine is composed by a bipartite quantum system, where one
subsystem is called visible, and the other is called hidden~\cite{RN3}. The type
of Hamiltonian of the bipartite quantum system is
\begin{equation}
  \label{eq:5}
  H(\mathbf{a}) = - \mathbf{a} \cdot \mathbf{O},
\end{equation}
where $\mathbf{a}$ is the vector of variable real parameters, and $\mathbf{O}$
is the vector of linearly independent Hermitian operators with zero trace. The
quantum state of the bipartite quantum system is the Boltzmann thermal
equilibrium state~\cite{RN4}
\begin{equation}
  \label{eq:1}
  \rho(\mathbf{a}) = \frac{e^{- H(\mathbf{a})}}{\Tr(e^{- H(\mathbf{a})})} =
  \frac{e^{\mathbf{a} \cdot \mathbf{O}}}{\Tr(e^{\mathbf{a} \cdot
      \mathbf{O}})}.
\end{equation}
Then the reduced state of the visible subsystem is
\begin{equation}
  \label{eq:2}
  \sigma(\mathbf{a}) = \Tr_{h}(\rho(\mathbf{a})),
\end{equation}
where $\Tr_{h}$ denotes the trace over the hidden subsystem.

The task of the quantum Boltzmann machine is to make the quantum state of the
visible subsystem $\sigma(\mathbf{a})$ approximate a given target state
$\sigma_{\ast}$ as well as possible by adjusting the parameters $\mathbf{a}$ in the
Hamiltonian. More precisely, the aim of the quantum Boltzmann machine is to
minimize
\begin{equation}
  \label{eq:102}
  S_{m}(\sigma_{\ast}) = \min_{\mathbf{a}} S(\sigma_{\ast}|\sigma(\mathbf{a})),
\end{equation}
where the quantum relative entropy is defined as
\begin{equation}
  \label{eq:4}
  S(\sigma|\sigma^{\prime}) = \Tr_{v}(\sigma(\ln \sigma - \ln \sigma^{\prime})).
\end{equation}
Here the quantum relative entropy is used as a measure of the degree
of approximation of one quantum state with another~\cite{Vedral02, RN8}.

From the above definition of quantum Boltzmann machine, we learn that a quantum
Boltzmann machine is specified by a set of operators $\mathcal{O}=\{O_{i}\}$,
where $\{O_{i}\}$ are linearly independent Hermitian operators. Then the set of
the Hamiltonians in Eq.~\eqref{eq:5} constructs a real linear space
$\mathcal{L}(\mathcal{O})$. Two quantum Boltzmann machines are of the same type
if and only if they have the same $\mathcal{L}(\mathcal{O})$. For a Boltzmann
machine with $\mathcal{L}(\mathcal{O})$, all the representable states are given
by Eqs.~\eqref{eq:1}\eqref{eq:2} denoted as $\mathcal{S}(\mathcal{O})$.

For convenience, we assume that any operator $O_{i}\in \mathcal{O}$ has a product
structure for the visible part and the hidden part, i.e.
$O_{i}=O_{i}^{v}\otimes O_{i}^{h}$, where $O_{i}^{v}$ and $O_{i}^{h}$ are the
Hermitian operators acting on the Hilbert space of the visible subsystem
$\mathcal{H}_{v}$ and that of the hidden subsystem $\mathcal{H}_{h}$
respectively. Then we divide the set $\mathcal{O}$ into three parts as follows:
\begin{equation}
  \label{eq:3}
  \mathcal{O} = \mathcal{O}_{v} \cup \mathcal{O}_{h} \cup \mathcal{O}_{c},
\end{equation}
where
\begin{align}
  \label{eq:10}
  \mathcal{O}_{v} & = \{O_{i}\in \mathcal{O}: O_{i}^{v} \neq 1^{v}, O_{i}^{h} = 1^{h}\}, \\
  \mathcal{O}_{h} & = \{O_{i}\in \mathcal{O}: O_{i}^{v} = 1^{v}, O_{i}^{h} \neq 1^{h}\}, \\
  \mathcal{O}_{c} & = \{O_{i}\in \mathcal{O}: O_{i}^{v} \neq 1^{v}, O_{i}^{h} \neq 1^{h}\},
\end{align}
with $1^{v}$ and $1^{h}$ being the identity operators of the Hilbert spaces
$\mathcal{H}_{v}$ and $\mathcal{H}_{h}$ respectively. The real linear spaces of
Hamiltonians with bases from $\mathcal{O}_{v}$, $\mathcal{O}_{h}$, and
$\mathcal{O}_{c}$ are denoted as $\mathcal{L}(\mathcal{O}_{h})$,
$\mathcal{L}(\mathcal{O}_{h})$, and $\mathcal{L}(\mathcal{O}_{c})$ respectively.
The linear space $\mathcal{L}(O)$ is the direct sum of the linear subspaces
$\mathcal{L}(\mathcal{O}_{v})$, $\mathcal{L}(\mathcal{O}_{h})$, and
$\mathcal{L}(\mathcal{O}_{c})$. Physically, a Hamiltonian in the composite
system can be naturally distinguished as the sum of the free Hamiltonian of the
visable subsystem, the free Hamiltonian of the hidden subsystem, and the
coupling term between them.

\section{Symmetry in quantum Boltzmann machine~\label{sec:symm-quant-boltzm}}

What is the symmetry for a quantum Boltzmann machine? Obviously, the concept of
the symmetry for a quantum Boltzmann machine is different from that for a given
quantum system. For a given quantum system, its Hamiltonian is fixed, and the
symmetric group is composed by the unitary transformations that keeps the
Hamiltonian invariant~\cite{Wigner}~\footnote{Here we do not consider the case
  where the symmetry group contains some anti-unitary transformations.}. For a
quantum Boltzmann machine, the Hamiltonian may be any one in the set of the real
linear space $\mathcal{L}(\mathcal{O})$. Thus we need to generalize the concept
of symmetry for a given Hamiltonian. In particular, to define the concept of the
symmetry for a quantum Boltzmann machine, the different roles played by the
visible part and the hidden part of the quantum Boltzmann machine must be taken
into account. Thus we introduce the symmetry of a quantum Boltzmann machine as
follows.

For the quantum Boltzmann machine with the Hamiltonian given by
Eq.~\eqref{eq:5}, we define its symmetry operation by the unitary transformation
$U_{v}\otimes U_{h}$ satisfying $\forall H(\mathbf{a}) \in \mathcal{L}(\mathcal{O})$
\begin{equation}
  \label{eq:6}
  U_{v}\otimes U_{h} H(\mathbf{a}) U_{v}^{\dagger}\otimes U_{h}^{\dagger} = H(\mathbf{a}^{\prime})\in
  \mathcal{L}(\mathcal{O}).
\end{equation}
All the symmetric unitary transformations form a group $G(\mathcal{O})$, which
is called the symmetry group of the quantum Boltzmann machine specified by
$\mathcal{O}$. In particular, the inverse of
$U_{v}\otimes U_{h}\in G(\mathcal{O})$ transforms $H(\mathbf{a}^{\prime})$ back to
$H(\mathbf{a})$, which implies that the map from $H(\mathbf{a})$ to
$H(\mathbf{a}^{\prime})$ is bijective. Further more, Eq.~\eqref{eq:6} gives
\begin{equation}
  \label{eq:9}
  U_{v}\otimes U_{h} \rho(\mathbf{a}) U_{v}^{\dagger}\otimes U_{h}^{\dagger} =
  \rho(\mathbf{a}^{\prime}),
\end{equation}
where $\rho(\mathbf{a})$ is defined by Eq.~\eqref{eq:1}.

For any symmetry unitary transformation $U_{v}\otimes U_{h}\in G(\mathcal{O})$ and any
target state $\sigma_{\ast}$, the transformed target state
$U_{v}\sigma_{\ast}U_{v}^{\dagger}$ will be approximated by the Boltzmann machine as well as
the original target state $\sigma_{\ast}$, more precisely,
\begin{equation}
  \label{eq:7}
  S_{m}(U_{v}\sigma_{\ast}U_{v}^{\dagger}) = S_{m}(\sigma_{\ast}),
\end{equation}
where $S_{m}(\sigma_{\ast})$ is defined by Eq.~\eqref{eq:102}. We prove Eq.~\eqref{eq:7}
as follows:
\begin{align}
  \label{eq:8}
  S_{m}(\sigma_{\ast}) & = \min_{\mathbf{a}} S(\sigma_{\ast}|\sigma(\mathbf{a}))
                         \nonumber\\
                       & = \min_{\mathbf{a}} S(U_{v} \sigma_{\ast} U_{v}^{\dagger}| U_{v} \sigma(\mathbf{a}) U_{v}^{\dagger})
                         \nonumber\\
                       & = \min_{\mathbf{a}} S(U_{v} \sigma_{\ast} U_{v}^{\dagger}| U_{v} \Tr_{h} \rho(\mathbf{a}) U_{v}^{\dagger})
                         \nonumber\\
                       & = \min_{\mathbf{a}} S(U_{v} \sigma_{\ast} U_{v}^{\dagger}| U_{v}
                         \Tr_{h}(1_{v}\otimes U_{h} \rho(\mathbf{a}) 1_{v}\otimes U_{h}^{\dagger}) U_{v}^{\dagger})
                         \nonumber\\
                       & = \min_{\mathbf{a}} S(U_{v} \sigma_{\ast} U_{v}^{\dagger}|
                         \Tr_{h}(U_{v}\otimes U_{h} \rho(\mathbf{a}) U_{v}^{\dagger} \otimes U_{h}^{\dagger}))
                         \nonumber\\
                       & = \min_{\mathbf{a}^{\prime}} S(U_{v} \sigma_{\ast} U_{v}^{\dagger}|
                         \Tr_{h}( \rho(\mathbf{a}^{\prime})))
                         \nonumber\\
                       & = S_{m}(U_{v} \sigma_{\ast} U_{v}^{\dagger}),
\end{align}
where the fact that the quantum relative entropy is invariant under unitary
transformation~\cite{RN12, Vedral02} is used on the second line, and
Eq.~\eqref{eq:9} is used on the fifth line. Eq.~\eqref{eq:7} shows that all the
target states related by the symmetry transformation $U_{v}$ of a quantum
Boltzmann machine are equivalent to the quantum Boltzmann machine.

Further more, we study the effect of the symmetry of quantum Boltzmann machine
on the possible solutions of parameters for a given target state. Let us denote
\begin{equation}
  \label{eq:54}
  \mathbf{A} = \argmin_{\mathbf{a}} S(\sigma_{\ast}|\sigma(\mathbf{a})),
\end{equation}
where the argmin of a function with some argments outputs the set of the
arguments where the function takes its minimum. If some
$\mathbf{a}_{\ast}\in \mathbf{A}$, then
$\forall U_{v}\otimes U_{h}\in G(\mathcal{O})$ satisfying $U_{v}\sigma_{\ast}U_{v}^{\dagger}=\sigma_{\ast}$, we have
\begin{equation}
  \label{eq:55}
  U_{v}\otimes U_{h} H(\mathbf{a}_{\ast}) U_{v}^{\dagger}\otimes U_{h}^{\dagger} = H(\mathbf{a}_{\ast}^{\prime}).
\end{equation}
Thus we prove that $\mathbf{a}_{\ast}^{\prime}\in\mathbf{A}$ as
follows:
\begin{align}
  \label{eq:56}
  S(\sigma_{\ast}|\sigma(\mathbf{a}_{\ast})) & = S(U_{v} \sigma_{\ast} U_{v}^{\dagger}|
                                               \Tr_{h}(U_{v}\otimes U_{h} \rho(\mathbf{a}_{\ast}) U_{v}^{\dagger} \otimes
                                               U_{h}^{\dagger})) \nonumber\\
                                             & = S(\sigma_{\ast}|\sigma(\mathbf{a}_{\ast}^{\prime})).
\end{align}
Roughly speaking, the symmetry of quantum Boltzmann machine implies
the degeneracy of the optimal solutions for a given target state.

Let us analyze the structure of the group $G(\mathcal{O})$. When the Hamiltonian
$H(\mathbf{a})$ in Eq.~\eqref{eq:6} is the free term of the visable subsystem,
i.e. $H(\mathbf{a})\in \mathcal{L}(\mathcal{O}_{v})$, Eq.~\eqref{eq:6} becomes
\begin{equation}
  \label{eq:11}
  U_{v} H(\mathbf{a}) U_{v}^{\dagger} = H(\mathbf{a}^{\prime}).
\end{equation}
All the unitary transformations satisfying the above equation construct a group,
denoted as $G(\mathcal{O}_{v})$. Similarly, we can introduce the group of the
hidden subsystem, denoted as $G(\mathcal{O}_{h})$. Then we can construct a
product group of $G(\mathcal{O}_{v})$ and $G(\mathcal{O}_{h})$, which contains
$G(\mathcal{O})$ as a subgroup. In fact, we can construct $G(\mathcal{O})$ by
\begin{equation}
  \label{eq:12}
  G(\mathcal{O}) = \{ U_{v}\otimes U_{h} \in G(\mathcal{O}_{v})\otimes
  G(\mathcal{O}_{h}): U_{v}\otimes U_{h} O U_{v}^{\dagger}\otimes
  U_{h}^{\dagger} \in \mathcal{L}(\mathcal{O}_{c}), \forall O\in \mathcal{L}(\mathcal{O}_{c})\}.
\end{equation}
In particular, when a quantum Boltzmann machine without the hidden subsystem,
the group $G(\mathcal{O})$ is the group $G(\mathcal{O}_{v})$. Further more,
Eq.~\eqref{eq:12} suggests that, to construct the symmetry group
$G(\mathcal{O})$ we can always construct the subgroups $G(\mathcal{O}_{v})$ and
$G(\mathcal{O}_{h})$ and then check the connect condition shown in
Eq.~\eqref{eq:12}.

\section{Quantum Boltzmann machine composed by qubits}
\label{sec:quant-boltzm-mach}

\subsection{Notation for multi-qubit systems}
\label{sec:notation-multi-qubit}

Usually a quantum Boltzmann machine is built on $n$ qubits, where the visible
part contains $n_{v}$ qubits, and the hidden part contains the other $n_{h}$
($n_{h}=n-n_{v}$) qubits. Let $Z_{2}=\{0,1\}$, which is an Abelian group under
$\oplus$, the addition modulo $2$. Let $\gamma=(\alpha,\beta)\in D$, where $D=Z_{2}\otimes Z_{2}$. Let
\begin{equation*}
  \boldsymbol{\gamma}=(\gamma_{1};\gamma_{2};\cdots;\gamma_{n})=(\alpha_{1},\beta_{1};\alpha_{2},\beta_{2};\cdots;\alpha_{n},\beta_{n})\in
  D^{n}.
\end{equation*}
The set of the $n$-qubit Pauli operators $\mathcal{P}_{n}$ is defined
by
\begin{equation}
  \label{eq:13}
  \mathcal{P}_{n} = \{ \sigma_{\boldsymbol{\gamma}}\equiv \sigma_{\gamma_{1}} \otimes \sigma_{\gamma_{2}} \otimes \cdots \otimes
  \sigma_{\gamma_{n}}\}
\end{equation}
with
\begin{equation}
  \label{eq:26}
  \sigma_{\gamma} = \sigma_{\alpha\beta} = i^{\alpha\beta} \sigma_{x}^{\alpha} \sigma_{z}^{\beta},
\end{equation}
where $\sigma_{x}$ and $\sigma_{z}$ are the Pauli operators~\cite{RN19}. Note that the set
$\mathcal{P}_{n}$ is a group up to a phase factor $\{\pm 1, \pm i\}$, which is
called the $n$-qubit Pauli group~\cite{RN12}. The advantage of the above
notation lies in the unified multiplication rule:
\begin{equation}
  \label{eq:27}
  \sigma_{\gamma}  \sigma_{\gamma^{'}}  = i^{\omega(\gamma,\gamma^{'})-\nu(\gamma,\gamma^{'})} \sigma_{\gamma\oplus\gamma^{'}}
\end{equation}
with
\begin{align}
  \label{eq:28}
  \omega(\gamma,\gamma^{'}) & = (\alpha+\alpha^{'})(\beta+\beta^{\prime}) - (\alpha\oplus\alpha^{'})(\beta\oplus\beta^{\prime}), \\
  \nu(\gamma,\gamma^{'}) & = \alpha\beta^{'}-\alpha^{'}\beta.
\end{align}
We observe that $\omega(\gamma,\gamma^{\prime})=\omega(\gamma^{\prime},\gamma)$ and
$\nu(\gamma,\gamma^{\prime})=-\nu(\gamma^{'},\gamma)$. In addition, the multiplication rule implies
\begin{align}
  \label{eq:34}
  \sigma_{\gamma}^{2} & = 1, \\
  \sigma_{\gamma} \sigma_{\gamma^{'}} & = (-1)^{\nu(\gamma,\gamma^{'})} \sigma_{\gamma^{'}}
                                        \sigma_{\gamma}.
\end{align}

In the $n$-qubit case, the multiplication rule becomes
\begin{equation}
  \label{eq:35}
  \sigma_{\boldsymbol{\gamma}}
  \sigma_{\boldsymbol{\gamma}^{'}} =i^{ \omega(\boldsymbol{\gamma},\boldsymbol{\gamma}^{\prime})-
    \nu(\boldsymbol{\gamma},\boldsymbol{\gamma}^{\prime})}
  \sigma_{\boldsymbol{\gamma}\oplus\boldsymbol{\gamma}^{'}},
\end{equation}
where
\begin{align}
  \label{eq:47}
  \omega(\boldsymbol{\gamma},\boldsymbol{\gamma}^{\prime}) & = \sum_{i} \omega(\gamma_{i},\gamma^{'}_{i}),
  \\
  \label{eq:48}
  \nu(\boldsymbol{\gamma},\boldsymbol{\gamma}^{\prime}) & = \sum_{i} \nu(\gamma_{i},\gamma^{'}_{i}).
\end{align}
And we also have
\begin{align}
  \label{eq:29}
  \sigma_{\boldsymbol{\gamma}}^{2} & = 1,\\
  \label{eq:53}
  \sigma_{\boldsymbol{\gamma}}
  \sigma_{\boldsymbol{\gamma}^{'}} & =(-1)^{\nu(\boldsymbol{\gamma},\boldsymbol{\gamma}^{\prime})}
                                     \sigma_{\boldsymbol{\gamma}^{'}} \sigma_{\boldsymbol{\gamma}}.
\end{align}

Since the elements in $\mathcal{P}_{n}$ are Hermitian and linear independent,
all the Hermitian operators can be expanded uniquely with them as a basis, which
form a real vector space
\begin{equation}
  \label{eq:30}
  \mathcal{L}_{n} = \left\{ \sum_{\boldsymbol{\gamma}} a_{\boldsymbol{\gamma}}
    \sigma_{\boldsymbol{\gamma}}, \boldsymbol{\gamma}\in D^{n},
    a_{\boldsymbol{\gamma}}\in \mathbf{R}\right\}.
\end{equation}

\subsection{Basic equations for the symmetry group of quantum Boltzmann machine}
\label{sec:basic-equat-symm}

The type of the quantum Boltzmann machine $\mathcal{O}$ is given by a subset of
$\mathcal{P}_{n}$, which is specified by a subset of $D^{n}$ denoted by
$D^{n}(\mathcal{O})$. The Hamiltonian space of the quantum Boltzmann machine
\begin{equation}
  \label{eq:31}
  \mathcal{L}(\mathcal{O})  = \left\{ \sum_{\boldsymbol{\gamma}}
    b_{\boldsymbol{\gamma}} \sigma_{\boldsymbol{\gamma}}, \boldsymbol{\gamma}\in D^{n}(\mathcal{O}),
    b_{\boldsymbol{\gamma}}\in \mathbf{R}\right\}.
\end{equation}
Similarly we can obtain its subspaces $\mathcal{L}(\mathcal{O}_{v})$
and $\mathcal{L}(\mathcal{O}_{h})$.

Now we start to construct the subgroup $G(\mathcal{O}_{v})$. Following
Eq.~\eqref{eq:11}, $\forall\boldsymbol{\gamma}\in D(\mathcal{O}_{v})$ and
$\forall U\in G(\mathcal{O}_{v})$ we have
\begin{equation}
  \label{eq:36}
  U \sigma_{\boldsymbol{\gamma}} U^{\dagger} = \sum_{\boldsymbol{\gamma}^{\prime}\in D(\mathcal{O}_{v})}
  \mathcal{U}_{\boldsymbol{\gamma} \boldsymbol{\gamma}^{\prime}} \sigma_{\boldsymbol{\gamma}^{\prime}},
\end{equation}
where
\begin{equation}
  \label{eq:49}
  \mathcal{U}_{\boldsymbol{\gamma} \boldsymbol{\gamma}^{\prime}} =
  \frac{1}{2^{n_{v}}} \Tr(U \sigma_{\boldsymbol{\gamma}} U^{\dagger} \sigma_{\boldsymbol{\gamma}^{\prime}}).
\end{equation}

Note that $\{\mathcal{U}_{\boldsymbol{\gamma}\boldsymbol{\gamma}^{\prime}}\}$ are real, and
they satisfy
\begin{equation}
  \label{eq:50}
  \sum_{\boldsymbol{\gamma}^{\prime}\in D(\mathcal{O}_{v})}
  \mathcal{U}_{\boldsymbol{\gamma}\boldsymbol{\gamma}^{\prime}}^{2} =
  \sum_{\boldsymbol{\gamma}\in D(\mathcal{O}_{v})}
  \mathcal{U}_{\boldsymbol{\gamma}\boldsymbol{\gamma}^{\prime}}^{2} = 1,
\end{equation}
which implies that
$\left(\mathcal{U}_{\boldsymbol{\gamma}\boldsymbol{\gamma}^{\prime}}^{2}\right)$ is a
double stochastic matrix~\cite{Bahtia1997,RN11}.
% We observe that $\mathcal{U}$ is a $n_{v}$-dimensional representation of $U$.
According to the Birkhoff's theorem~\cite{Bahtia1997}, the set of all
$n_{v}\times n_{v}$ doubly stochastic matrices is a convex set whose extreme points
are permutation matrices.

According to Eq.~\eqref{eq:35}, we define the generator $D_{a}(\mathcal{O}_{v})$
of $D(\mathcal{O}_{v})$ under the addition modulo $2$ as a subset of
$D(\mathcal{O}_{v})$ such that
$\forall \boldsymbol{\gamma}\in D(\mathcal{O}_{v})$ there exists a unique decomposition
\begin{equation}
  \label{eq:42}
  \boldsymbol{\gamma} = \oplus_{\boldsymbol{\gamma}^{\prime}\in D_{a}(\mathcal{O}_{v})}
  \boldsymbol{\gamma}^{\prime},
\end{equation}
which implies that
\begin{equation}
  \label{eq:52}
  \sigma_{\boldsymbol{\gamma}} = c_{\boldsymbol{\gamma}} \otimes_{\boldsymbol{\gamma}^{\prime}\in
    D_{a}(\mathcal{O}_{v})} \sigma_{\boldsymbol{\gamma}^{\prime}}
\end{equation}
with $c_{\boldsymbol{\gamma}}\in\{\pm 1,\pm i\}$.

Thus the independent relations in Eq.~\eqref{eq:36} are given by
$\forall \boldsymbol{\gamma}\in D_{a}(\mathcal{O}_{v})$
\begin{equation}
  \label{eq:43}
  U \sigma_{\boldsymbol{\gamma}} U^{\dagger} = \sum_{\boldsymbol{\gamma}^{\prime}\in
    D(\mathcal{O}_{v})} \mathcal{U}_{\boldsymbol{\gamma}
    \boldsymbol{\gamma}^{\prime}} \sigma_{\boldsymbol{\gamma}^{\prime}}.
\end{equation}
Applying $U$ to Eq.~\eqref{eq:29} gives
$\forall \boldsymbol{\gamma}\in D_{a}(\mathcal{O}_{v})$,
\begin{align}
  \label{eq:44}
  \left( U \sigma_{\boldsymbol{\gamma}} U^{\dagger} \right)^{2}
  & = \sum_{\boldsymbol{\gamma}^{\prime\prime},\boldsymbol{\gamma}^{\prime\prime\prime}\in
    D(\mathcal{O}_{v})} \mathcal{U}_{\boldsymbol{\gamma}
    \boldsymbol{\gamma}^{\prime\prime}} \mathcal{U}_{\boldsymbol{\gamma}
    \boldsymbol{\gamma}^{\prime\prime\prime}} \sigma_{\boldsymbol{\gamma}^{\prime\prime}}
    \sigma_{\boldsymbol{\gamma}^{\prime\prime\prime}}
    \nonumber\\
  & = \sum_{\boldsymbol{\gamma}^{\prime\prime}\in
    D(\mathcal{O}_{v})} \mathcal{U}_{\boldsymbol{\gamma}
    \boldsymbol{\gamma}^{\prime\prime}}^{2}
    + \sum_{\boldsymbol{\gamma}^{\prime\prime}\neq\boldsymbol{\gamma}^{\prime\prime\prime}\in
    D(\mathcal{O}_{v})} \mathcal{U}_{\boldsymbol{\gamma}
    \boldsymbol{\gamma}^{\prime\prime}} \mathcal{U}_{\boldsymbol{\gamma}
    \boldsymbol{\gamma}^{\prime\prime\prime}} \sigma_{\boldsymbol{\gamma}^{\prime\prime}}
    \sigma_{\boldsymbol{\gamma}^{\prime\prime\prime}} \nonumber\\
  & = \sum_{\boldsymbol{\gamma}^{\prime\prime}\in
    D(\mathcal{O}_{v})} \mathcal{U}_{\boldsymbol{\gamma}
    \boldsymbol{\gamma}^{\prime\prime}}^{2}
    + \sum_{\boldsymbol{\gamma}^{\prime\prime}\neq\boldsymbol{\gamma}^{\prime\prime\prime}\in
    D(\mathcal{O}_{v})} \mathcal{U}_{\boldsymbol{\gamma}
    \boldsymbol{\gamma}^{\prime\prime}} \mathcal{U}_{\boldsymbol{\gamma}
    \boldsymbol{\gamma}^{\prime\prime\prime}}
    i^{\omega(\boldsymbol{\gamma}^{\prime\prime},\boldsymbol{\gamma}^{\prime\prime\prime})
    -
    \nu(\boldsymbol{\gamma}^{\prime\prime},\boldsymbol{\gamma}^{\prime\prime\prime})}
    \sigma_{\boldsymbol{\gamma}^{\prime\prime}\oplus\boldsymbol{\gamma}^{\prime\prime\prime}}
    \nonumber\\
  & = 1.
\end{align}
Thus we find
$\forall \boldsymbol{\gamma}_{\ast} \in D^{n_{v}}$ and
$\forall \boldsymbol{\gamma}\in D_{a}(\mathcal{O}_{v})$,
\begin{align}
  \label{eq:45}
  \sum_{\boldsymbol{\gamma}^{\prime\prime}\in
  D(\mathcal{O}_{v})}
  \mathcal{U}_{\boldsymbol{\gamma}\boldsymbol{\gamma}^{\prime\prime}}^{2} & = 1, \\
  \label{eq:17}
  \sum_{\boldsymbol{\gamma}^{\prime\prime}\neq\boldsymbol{\gamma}^{\prime\prime\prime}\in
  D(\mathcal{O}_{v})} \mathcal{U}_{\boldsymbol{\gamma}
  \boldsymbol{\gamma}^{\prime\prime}} \mathcal{U}_{\boldsymbol{\gamma}
  \boldsymbol{\gamma}^{\prime\prime\prime}}
  i^{\omega(\boldsymbol{\gamma}^{\prime\prime},\boldsymbol{\gamma}^{\prime\prime\prime})
  -
  \nu(\boldsymbol{\gamma}^{\prime\prime},\boldsymbol{\gamma}^{\prime\prime\prime})}
  \delta_{\boldsymbol{\gamma}^{\ast},\boldsymbol{\gamma}^{\prime\prime}\oplus\boldsymbol{\gamma}^{\prime\prime\prime}} & = 0.
\end{align}

Similarly, applying $U$ to Eq.~\eqref{eq:53}, we find
$\forall \boldsymbol{\gamma}_{\ast} \in D^{n_{v}}$ and
$\forall \boldsymbol{\gamma}\neq \boldsymbol{\gamma}^{\prime}\in D_{a}(\mathcal{O}_{v})$,
\begin{align}
  \label{eq:46}
  \sum_{\boldsymbol{\gamma}^{\prime\prime}\neq\boldsymbol{\gamma}^{\prime\prime\prime}\in
  D(\mathcal{O}_{v})} \left( \mathcal{U}_{\boldsymbol{\gamma}
  \boldsymbol{\gamma}^{\prime\prime}} \mathcal{U}_{\boldsymbol{\gamma}^{\prime}
  \boldsymbol{\gamma}^{\prime\prime\prime}} - (-1)^{\nu(\boldsymbol{\gamma},\boldsymbol{\gamma}^{\prime})}\mathcal{U}_{\boldsymbol{\gamma}^{\prime}
  \boldsymbol{\gamma}^{\prime\prime}} \mathcal{U}_{\boldsymbol{\gamma}
  \boldsymbol{\gamma}^{\prime\prime\prime}} \right)
  i^{\omega(\boldsymbol{\gamma}^{\prime\prime},\boldsymbol{\gamma}^{\prime\prime\prime})
  -
  \nu(\boldsymbol{\gamma}^{\prime\prime},\boldsymbol{\gamma}^{\prime\prime\prime})}
  \delta_{\boldsymbol{\gamma}^{\ast},\boldsymbol{\gamma}^{\prime\prime}\oplus\boldsymbol{\gamma}^{\prime\prime\prime}} & = 0.
\end{align}

Following Eq.~\eqref{eq:52}, $\forall \boldsymbol{\gamma}\in D(\mathcal{O}_{v})$ and
$\boldsymbol{\gamma}\notin D_{a}(\mathcal{O}_{v})$, we require
\begin{align}
  \label{eq:14}
  U \sigma_{\boldsymbol{\gamma}} U^{\dagger}
  & = c_{\gamma} \otimes_{\boldsymbol{\gamma}^{\prime}} U \sigma_{\boldsymbol{\gamma}^{\prime}} U^{\dagger}
    \nonumber\\
  & = c_{\gamma} \otimes_{\boldsymbol{\gamma}^{\prime}} \sum_{\boldsymbol{\gamma}^{\prime\prime}}
    \mathcal{U}_{\boldsymbol{\gamma}^{\prime}\boldsymbol{\gamma}^{\prime\prime}} \sigma_{\boldsymbol{\gamma}^{\prime\prime}}
    \in \mathcal{L}(\mathcal{O}_{v}).
\end{align}
For example, if the decomposition of $\sigma_{\boldsymbol{\gamma}}$
contains two terms $\sigma_{\boldsymbol{\gamma}_{1}^{\prime}}$ and
$\sigma_{\boldsymbol{\gamma}_{2}^{\prime}}$, then Eq.~\eqref{eq:14}
becomes
\begin{align}
  \label{eq:15}
  U \sigma_{\boldsymbol{\gamma}} U^{\dagger}
  & = c_{\gamma}  \sum_{\boldsymbol{\gamma}_{1}^{\prime\prime}, \boldsymbol{\gamma}_{2}^{\prime\prime}}
    \mathcal{U}_{\boldsymbol{\gamma}_{1}^{\prime}\boldsymbol{\gamma}_{1}^{\prime\prime}}
    \mathcal{U}_{\boldsymbol{\gamma}_{2}^{\prime}\boldsymbol{\gamma}_{2}^{\prime\prime}}
    \sigma_{\boldsymbol{\gamma}_{1}^{\prime\prime}}  \sigma_{\boldsymbol{\gamma}_{2}^{\prime\prime}}
    \nonumber\\
  & = c_{\gamma}  \sum_{\boldsymbol{\gamma}_{1}^{\prime\prime}, \boldsymbol{\gamma}_{2}^{\prime\prime}}
    \mathcal{U}_{\boldsymbol{\gamma}_{1}^{\prime}\boldsymbol{\gamma}_{1}^{\prime\prime}}
    \mathcal{U}_{\boldsymbol{\gamma}_{2}^{\prime}\boldsymbol{\gamma}_{2}^{\prime\prime}}
    i^{\omega(\boldsymbol{\gamma}_{1}^{\prime\prime},\boldsymbol{\gamma}_{2}^{\prime\prime}) -
    \nu(\boldsymbol{\gamma}_{1}^{\prime\prime},\boldsymbol{\gamma}_{2}^{\prime\prime})}
    \sigma_{\boldsymbol{\gamma}_{1}^{\prime\prime}\oplus \boldsymbol{\gamma}_{2}^{\prime\prime}} \in
    \mathcal{L}(\mathcal{O}_{v}),
\end{align}
which implies that
$\forall \mathcal{\gamma}_{\ast} \notin D(\mathcal{O}_{v})$,
\begin{equation}
  \label{eq:16}
  \sum_{\boldsymbol{\gamma}_{1}^{\prime\prime}, \boldsymbol{\gamma}_{2}^{\prime\prime}}
  \mathcal{U}_{\boldsymbol{\gamma}_{1}^{\prime}\boldsymbol{\gamma}_{1}^{\prime\prime}}
  \mathcal{U}_{\boldsymbol{\gamma}_{2}^{\prime}\boldsymbol{\gamma}_{2}^{\prime\prime}}
  i^{\omega(\boldsymbol{\gamma}_{1}^{\prime\prime},\boldsymbol{\gamma}_{2}^{\prime\prime}) -
    \nu(\boldsymbol{\gamma}_{1}^{\prime\prime},\boldsymbol{\gamma}_{2}^{\prime\prime})}
  \delta_{\mathcal{\gamma}_{\ast}, \boldsymbol{\gamma}_{1}^{\prime\prime}\oplus
    \boldsymbol{\gamma}_{2}^{\prime\prime}} = 0.
\end{equation}

Now we concludes that Eqs.~\eqref{eq:45}\eqref{eq:17}\eqref{eq:46}\eqref{eq:16}
construct the basic equations for the coefficients
$\{\mathcal{U}_{\boldsymbol{\gamma} \boldsymbol{\gamma}^{\prime}},
\boldsymbol{\gamma}\in D_{a}(\mathcal{O}_{v}), \boldsymbol{\gamma}^{\prime}\in
D(\mathcal{O}_{v})\}$. Actually, our basic equations ensure that the
transformation $U$ is unitary, and that the set $\mathcal{O}_{v}$ can be
generated with the new generator with the same rule as the original. Thus, in
principle, all the elements in the group $G(\mathcal{O}_{v})$ can be determined
by solving these basic equations.

\subsection{The continuous and the discrete subgroups of quantum Bolztmann
  machine }
\label{sec:cont-discr-subgr}

Since the basic equations for the symmetry group $G(\mathcal{O}_{v})$ for the
Boltzmann machine is difficult to solve directly, we will find the analytical
results on the symmetry group in this subsection. We note that the elements in
the group $G(\mathcal{O}_{v})$ may be divided into two classes, the continuous
part $G_{c}(\mathcal{O}_{v})$ and the discrete part $G_{d}(\mathcal{O}_{v})$.

Now we focus on the continuous subgroup $G_{c}(\mathcal{O}_{v})$. First, the
one-parameter subgroup~\cite{Wigner} of $G(\mathcal{O}_{v})$ may be written as
\begin{equation}
  \label{eq:23}
  U(a) = e^{i a K }
\end{equation}
with
\begin{equation}
  \label{eq:41}
  K = \sum_{\boldsymbol{\gamma}\in D^{n_{v}}} b_{\boldsymbol{\gamma}} \sigma_{\boldsymbol{\gamma}}.
\end{equation}

Then $\forall \boldsymbol{\gamma}^{\prime}\in G(\mathcal{O}_{v})$, we have
\begin{align}
  \label{eq:24}
  & \phantom{=}  \comm{K}{ \sigma_{\boldsymbol{\gamma}^{\prime}}}
    \nonumber\\
  & = \comm{\sum_{\boldsymbol{\gamma}\in D^{n_{v}}} b_{\boldsymbol{\gamma}}
    \sigma_{\boldsymbol{\gamma}}}{ \sigma_{\boldsymbol{\gamma}^{\prime}}}  \nonumber\\
  & = \sum_{\boldsymbol{\gamma}\in D^{n_{v}}} b_{\boldsymbol{\gamma}}
    \comm{\sigma_{\boldsymbol{\gamma}}}{\sigma_{\boldsymbol{\gamma}^{\prime}}} \nonumber\\
  & = \sum_{\boldsymbol{\gamma}\in D^{n_{v}}} b_{\boldsymbol{\gamma}} (1 -
    (-1)^{\nu(\boldsymbol{\gamma},\boldsymbol{\gamma}^{\prime})})
    i^{\omega(\boldsymbol{\gamma},\boldsymbol{\gamma}^{\prime})- \nu(\boldsymbol{\gamma},\boldsymbol{\gamma}^{\prime})}
    \sigma_{\boldsymbol{\gamma}\oplus\boldsymbol{\gamma}^{'}} \in
    \mathcal{L}(\mathcal{O}_{v}),
\end{align}
which implies that if
$\boldsymbol{\gamma}\oplus\boldsymbol{\gamma}^{\prime}\notin
D(\mathcal{O}_{v})$, then
\begin{equation}
  \label{eq:37}
  b_{\boldsymbol{\gamma}} (1 - (-1)^{\nu(\boldsymbol{\gamma},\boldsymbol{\gamma}^{\prime})}) = 0.
\end{equation}
In other words, $b_{\boldsymbol{\gamma}}\neq 0$ if and only if
$\forall \boldsymbol{\gamma}^{\prime}\in D(\mathcal{O}_{v})$
\begin{equation}
  \label{eq:38}
  \boldsymbol{\gamma}\oplus\boldsymbol{\gamma}^{\prime}\in D(\mathcal{O}_{v})
\end{equation}
or
\begin{equation}
  \label{eq:39}
  \comm{\sigma_{\boldsymbol{\gamma}}}{\sigma_{\boldsymbol{\gamma}^{\prime}}} = 0,
\end{equation}
which is equivalent to
\begin{equation}
  \label{eq:40}
  \nu(\boldsymbol{\gamma},\boldsymbol{\gamma}^{\prime}) \pmod 2 = 0.
\end{equation}
Let us denote the set of $\boldsymbol{\gamma}$ that satisfies Eq.~\eqref{eq:38}
or Eq.~\eqref{eq:40} for any
$\boldsymbol{\gamma}^{\prime}\in D(\mathcal{O}_{v})$ as $D_{c}$. Then
$U(\{a_{\boldsymbol{\gamma}}\})=e^{i\sum_{\boldsymbol{\gamma}\in D_{c}}
  a_{\boldsymbol{\gamma}} \sigma_{\boldsymbol{\gamma}}}$ is the continuous
subgroup of $G(\mathcal{O}_{v})$, which is denoted as $G_{c}(\mathcal{O}_{v})$.

We find that $G_{c}(\mathcal{O}_{v})$ is an invariant group of
$G(\mathcal{O}_{v})$, which is proved as follows:
$\forall V\in G(\mathcal{O}_{v})$, we have
\begin{align}
  \label{eq:25}
  V U(\{a_{\boldsymbol{\gamma}}\}) V^{\dagger}
  & = e^{i\sum_{\boldsymbol{\gamma}\in D_{c}} a_{\boldsymbol{\gamma}} V \sigma_{\boldsymbol{\gamma}} V^{\dagger}} \in
    G_{c}(\mathcal{O}_{v})\nonumber\\
  & = U(\{a^{'}_{\boldsymbol{\gamma}}\}).
\end{align}
Thus we conclude that the group $G(\mathcal{O}_{v})$ is a product of the
continuous subgroup $G_{c}(\mathcal{O}_{v})$ and some discrete subgroup
$G_{d}(\mathcal{O}_{v})$, i.e.,
$G(\mathcal{O}_{v})=G_{c}(\mathcal{O}_{v})\otimes G_{d}(\mathcal{O}_{v})$. In
other words, any element in $G(\mathcal{O}_{v})$ can be written as
$U=U(\{a_{\boldsymbol{\gamma}}\})W$ with $W\in G_{d}(\mathcal{O}_{v})$.

Now we focus on constructing the discrete subgroup $G_{d}(\mathcal{O}_{v})$. A
direct method is as follows. We numerically solve the set of basic
equations~\eqref{eq:45}\eqref{eq:17}\eqref{eq:46}\eqref{eq:16} with random
initial conditions to obtain a random element in $G(\mathcal{O}_{v})$. Combining
with the known $G_{c}(\mathcal{O}_{v})$, we will obtain a random element in
$G_{d}(\mathcal{O}_{v})$. Repeat the above procedure until all the elements in
$G_{d}(\mathcal{O}_{v})$ be found. However, this method only works when the
number of the qubit in the system is small (less than $5$ qubits).

To obtain the discrete group $G_{d}(\mathcal{O}_{v})$ in the systems with more
qubits, we make the assumption that $G_{d}(\mathcal{O}_{v})$ is a subgroup of
the Clifford group~\cite{RN13,RN14} of $n_{v}$ qubits, where the Clifford group
is defined as all the unitary transformations that map one element to another in
$\mathcal{P}_{n_{v}}$ up to the sign $\pm$. This assumption is numerically
supported for the lower dimensional system by the method given in the previous
paragraph.

Under the above assumption, Eqs.~\eqref{eq:45}\eqref{eq:17} in the basic
equations are satisfied automatically. Eqs.~\eqref{eq:46} ensures the
commutation relations between different elements in the generator of the set
$\mathcal{O}_{v}$ invariant after transformations. Since the commutation
relation between any two elements in the Pauli group is either commutative or
anti-commutative, we can describe these relations by a graph, where an vertex
denotes an element in the Pauli group, and an edge between two nodes denotes the
two elements in the Pauli group are anti-commutative. In the language of
graph~\cite{RN15}, the commutation relations among the elements in the generator
of the set $\mathcal{O}_{v}$ can be described as a graph, and a graph is
transformed to another isomorphic graph under the unitary transformation in the
group $G_{d}(\mathcal{O}_{v})$~\cite{RN16, RN17, RN18}. Eqs.~\eqref{eq:16}
ensures the set $\mathcal{O}_{v}$ can be generated with the new generator with
the same rule as original.

So far we are ready to present a general procedure to determine the symmetry
group of a quantum Boltzmann machine $\mathcal{O}$:
\begin{enumerate}
\item Give $D(\mathcal{O})$, and classify it into $D(\mathcal{O}_{v})$,
  $D(\mathcal{O}_{h})$, and $D(\mathcal{O}_{c})$.
\item\label{item:1} Calculate the subgroup $G(\mathcal{O}_{v})$.
  \begin{enumerate}
  \item Calculate the continuous subgroup $G_{c}(\mathcal{O}_{v})$ by
    Eqs.~\eqref{eq:38}\eqref{eq:40}.
  \item\label{item:4} Calculate the discrete subgroup
    $G_{d}(\mathcal{O}_{v})$. \\
    First, calculate the commutation relations
    $\nu(\boldsymbol{\gamma},\boldsymbol{\gamma})\;\mod 2$ in
    $D(\mathcal{O}_{v})$ and associate it with a graph. Second, select one
    generator $D_{a}(\mathcal{O}_{v})$ and obtain its graph. Find all the other
    generator $D_{a,k}(\mathcal{O}_{v})$ with the isomorphic graph. Third, find
    all the generators in $D_{a,k}(\mathcal{O}_{v})$ such that all the elements
    in $D(\mathcal{O}_{v})$ can be generated from $D_{a,k}(\mathcal{O}_{v})$
    with the same rule as from $D_{a}(\mathcal{O}_{v})$. Then any map from
    $D_{a}(\mathcal{O}_{v})$ to $D_{a,k}(\mathcal{O}_{v})$ defines an element in
    $G_{d}(O_{v})$.
  \end{enumerate}
\item\label{item:2} Calculate the subgroup $G(\mathcal{O}_{h})$ Similarly.
\item\label{item:3} Determine the group $G(\mathcal{O})$ by checking the connect
  condition in Eq.~\eqref{eq:12}.
\end{enumerate}

\subsection{Examples of constructing the symmetry group of quantum Boltzmann
  machine}
\label{sec:exampl-constr-symm}

\begin{figure}[htbp]
  \centering
  \begin{tikzpicture}
    \begin{scope}[every node/.style={draw, circle, fill=gray!20}]
      \node (n1) at (0, 0) {$\boldsymbol{\gamma}_{1}$};
      \node (n2) at (2, 0) {$\boldsymbol{\gamma}_{2}$};
      \node (n3) at (0, 2) {$\boldsymbol{\gamma}_{3}$};
      \node (n4) at (2, 2) {$\boldsymbol{\gamma}_{4}$};
      \node (n5) at (-2, 1) {$\boldsymbol{\gamma}_{5}$};
      \draw (n1) -- (n2) (n3) -- (n4) (n1) -- (n5)  (n3) -- (n5);
    \end{scope}
  \end{tikzpicture}
  \caption{A graph showing commutation relations in $\mathcal{O}_{I}$.}
  \label{fig:3}
\end{figure}
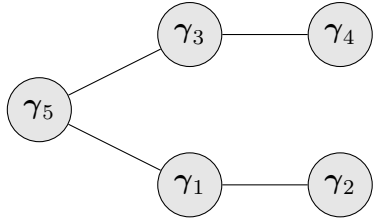

Let us demonstrate the above procedure with the quantum Boltzmann machine with
the Hamiltonian:
\begin{equation}
  H_{I}= a_1 \sigma_{x}^{(1)} + a_2 \sigma_{z}^{(1)} + a_3 \sigma_{x}^{(2)} + a_4
  \sigma_{z}^{(2)}  + a_5 \sigma_{z}^{(1)} \otimes \sigma_{z}^{(2)}.
\end{equation}
Then the set
\begin{equation}
  \label{eq:18}
  D(\mathcal{O}_{I}) = \{ \boldsymbol{\gamma}_{1} = (1,0;0,0),
  \boldsymbol{\gamma}_{2} = (0,1;0,0), \boldsymbol{\gamma}_{3} = (0,0;1,0),
  \boldsymbol{\gamma}_{4} = (0,0;0,1), \boldsymbol{\gamma}_{5} = (0,1;0,1).\}
\end{equation}
The commutation relations, which are characterized by the matrix
$\nu(\boldsymbol{\gamma}_{i}, \boldsymbol{\gamma}_{j}) \mod 2$, can be
described by the graph in Fig.~\ref{fig:3}.
To obtain the continuous subgroup $G_{c}(\mathcal{O}_{I})$, we need to solve
Eqs.~\eqref{eq:38}\eqref{eq:40}. We find that there is only one solution,
$\boldsymbol{\gamma}=(0,0;0,0)$, which means that $G_{c}(\mathcal{O}_{I})=I$.

\begin{figure}[htbp]
  \centering
  \begin{tikzpicture}
    \begin{scope}[every node/.style={draw, circle, fill=gray!20}]
      \node (n1) at (0, 0) {$\boldsymbol{\gamma}_{1}$};
      \node (n2) at (2, 0) {$\boldsymbol{\gamma}_{2}$};
      \node (n3) at (0, 2) {$\boldsymbol{\gamma}_{3}$};
      \node (n4) at (2, 2) {$\boldsymbol{\gamma}_{4}$};
      \draw (n1) -- (n2) (n3) -- (n4);
      \node (nn1) at (5, 0) {$\boldsymbol{\gamma}_{1}^{\prime}$};
      \node (nn2) at (7, 0) {$\boldsymbol{\gamma}_{2}^{\prime}$};
      \node (nn3) at (5, 2) {$\boldsymbol{\gamma}_{3}^{\prime}$};
      \node (nn4) at (7, 2) {$\boldsymbol{\gamma}_{4}^{\prime}$};
      \draw (nn1) -- (nn2) (nn3) -- (nn4);
    \end{scope}
    \draw[->] (2.5,1) to[above] node {$U$} (4.5,1);
  \end{tikzpicture}
  \caption[The unitary transformation in the discrete group
  $G_{d}(\mathcal{O}_{I})$]{Isomorphism under unitary transformation}
  \label{fig:2}
\end{figure}
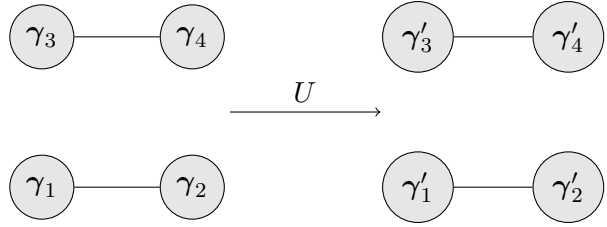

To study the discrete subgroup $G_{d}(\mathcal{O}_{I})$, let the generator of
$\mathcal{O}_{I}$ be
$D_{a}(\mathcal{O}_{v})=\{\boldsymbol{\gamma}_{1}, \boldsymbol{\gamma}_{2},
\boldsymbol{\gamma}_{3}, \boldsymbol{\gamma}_{4}\}$, whose commutation relations
are demonstrated in a graph in Fig.~\ref{fig:2}. Under the unitary
transformation $U\in G_{d}(\mathcal{O}_{I})$, the generator
$D_{a}(\mathcal{O}_{v})$ becomes $D_{a}(\mathcal{O}_{v}^{\prime})$ that
preserves the commutation relations.

We observe that only $\boldsymbol{\gamma}_{5}^{\prime}=\boldsymbol{\gamma}_{5}$
the remaining part in the graph in Fig.~\ref{fig:3} is isomorphic to the left
side of the graph in Fig.~\ref{fig:2}. Since
$\boldsymbol{\gamma}_{5}=\boldsymbol{\gamma}_{2}\oplus \boldsymbol{\gamma}_{4}$,
correspondingly we have
$\boldsymbol{\gamma}_{5}^{\prime}=\boldsymbol{\gamma}_{2}^{\prime}\oplus
\boldsymbol{\gamma}_{4}^{\prime}$. Thus we have two choices:
$\{\boldsymbol{\gamma}_{2}^{\prime}=\boldsymbol{\gamma}_{2},
\boldsymbol{\gamma}_{4}^{\prime}=\boldsymbol{\gamma}_{4}\}$ or
$\{\boldsymbol{\gamma}_{2}^{\prime}=\boldsymbol{\gamma}_{4},
\boldsymbol{\gamma}_{4}^{\prime}=\boldsymbol{\gamma}_{2}\}$. Together with the
commutation relations between $D_{a}(\mathcal{O}_{v}^{\prime})$, we finally
obtain
\begin{equation}
  \label{eq:19}
  \boldsymbol{\gamma}_{1}^{\prime}=\boldsymbol{\gamma}_{1},
  \boldsymbol{\gamma}_{2}^{\prime}=\boldsymbol{\gamma}_{2},
  \boldsymbol{\gamma}_{3}^{\prime}=\boldsymbol{\gamma}_{3},
  \boldsymbol{\gamma}_{4}^{\prime}=\boldsymbol{\gamma}_{4},
  \boldsymbol{\gamma}_{5}^{\prime}=\boldsymbol{\gamma}_{5};
\end{equation}
or
\begin{equation}
  \label{eq:20}
  \boldsymbol{\gamma}_{1}^{\prime}=\boldsymbol{\gamma}_{3},
  \boldsymbol{\gamma}_{2}^{\prime}=\boldsymbol{\gamma}_{4},
  \boldsymbol{\gamma}_{3}^{\prime}=\boldsymbol{\gamma}_{1},
  \boldsymbol{\gamma}_{4}^{\prime}=\boldsymbol{\gamma}_{2},
  \boldsymbol{\gamma}_{5}^{\prime}=\boldsymbol{\gamma}_{5}.
\end{equation}
Therefore the group $G_{d}(\mathcal{O}_{I})$ is
\begin{equation}
  G_{d}(\mathcal{O}_{I}) = \mathcal{S}^{S}_{1,2} \otimes  \mathcal{P}_{1,2},
\end{equation}
where $\mathcal{S}^{S}_{a_{1},a_{2},\cdots,a_{n}}$ is the $n$-qubit permutation
group containing all possible permutations of the qubits labeled from $a_{1}$ to
$a_{n}$. Here $\mathcal{S}^{S}_{1,2} = \{I, S_{12}\}$, and $S_{ij}$ is the swap
gate between qubit $i$ and qubit $j$. $\mathcal{P}_{a_{1},a_{2},\cdots,a_{n}}$
are the $n$-qubit Pauli group for the qubits labeled from $a_{1}$ to $a_{n}$ the
qubits labeled from $a_{1}$ to $a_{n}$.

Note that the above quantum Boltzmann machine can be extended to its $n$-qubit
version, where the Hamiltonian is
\begin{equation}
  H_{II} =  \sum_{i=1}^{n} \left(a_{2i-1} \sigma_{x}^{(i)} + a_{2i}
    \sigma_{z}^{(i)}\right) +  \sum_{i<j}^{n} a_{i,j} \sigma_{z}^{(i)} \otimes \sigma_{z}^{(j)}
\end{equation}
Similarly we find $G_{c}(\mathcal{O}_{II}) = I$, and the discrete
subgroup
\begin{equation}
\label{eq:32}
  G_{d}(\mathcal{O}_{II})= \mathcal{S}^{S}_{1,2,\cdots,n} \otimes  \mathcal{P}_{1,2,\cdots,n}.
\end{equation}
From Eq.~\eqref{eq:12} we know
\begin{equation}
\label{eq:33}
  G(\mathcal{O}_{II}) = G_c(\mathcal{O}_{II}) \otimes G_d(\mathcal{O}_{II}) = G_d(\mathcal{O}_{II}).
\end{equation}
	
\begin{figure}[htbp]
  \centering
  \begin{tikzpicture}
    \begin{scope}[every node/.style={draw, circle, fill=gray!20}]
      \node (n1) at (0, 0) {};
      \node (n2) at (2, 0) {};
      \node (n3) at (0, 2) {};
      \node (n4) at (2, 2) {};
      \node (n5) at (-2, 1) {};
      \node (n6) at (4, 1) {};
      \node (n7) at (1, 0.6) {};
      \node (n8) at (1, 1.4) {};
      \draw (n1) -- (n2) (n3) -- (n4) (n1) -- (n5)  (n3) -- (n5) (n2) -- (n6)
      (n4) -- (n6) (n5) -- (n7) (n6) -- (n7) (n5) -- (n8) (n6) -- (n8) (n2) --
      (n7) (n3) -- (n7) (n1) -- (n8) (n4) -- (n8);
    \end{scope}
  \end{tikzpicture}
  \caption{A graph showing commutation relations in $\mathcal{O}_{III}$ with
    $n=2$.}
  \label{fig:4}
\end{figure}
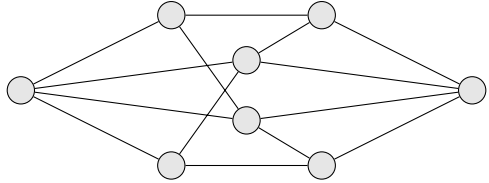

We now consider the quantum Boltzmann machine with the ZZXX-XZ Hamiltonian:
\begin{equation}
  H_{III} =  \sum_{i=1}^{n} \left(a_{2i-1} \sigma_{x}^{(i)} + a_{2i}
    \sigma_{z}^{(i)}\right) +  \sum_{i<j}^{n} \left(a_{i,j} \sigma_{z}^{(i)} \otimes
    \sigma_{z}^{(j)} + b_{i,j} \sigma_{x}^{(i)} \otimes \sigma_{x}^{(j)}\right) +
  \sum_{i\neq j}^{n} c_{i,j} \sigma_{x}^{(i)} \otimes \sigma_{z}^{(j)}.
\end{equation}
The commutation relations in $\mathcal{O}_{III}$ is demonstrated in
Fig.~\ref{fig:4}.

According to Eq.~\eqref{eq:38}\eqref{eq:40}, we obtain
\begin{equation}
  \sigma_{\boldsymbol{\gamma}_{j}} = \sigma_{y}^{(j)}.
\end{equation}
The continuous subgroup is
\begin{equation}
  U(\{a_{j}\}) = e^{i \sum_{j} a_{j} \sigma_{y}^{(j)}}.
\end{equation}
And the discrete subgroup is
\begin{equation}
\label{eq:51}
  G_{d}(\mathcal{O}_{III}) = \mathcal{S}^{H}_{n}
  \otimes \mathcal{S}^{S}_{1,2,\cdots,n} \otimes
  \mathcal{P}_{1,2,\cdots,n},
\end{equation}
where $\mathcal{S}^{H}_{n}$ is the $n$-qubit permutation group generating by generators $\{I, H_{i}\}$, 
and $H_{i}$ is the permutation between $\sigma_{x}^{(i)}$ and $\sigma_{z}^{(i)}$.

Because $\mathcal{S}^{H}_{n} \subset U(\{a_{j}\})$, from the uniqueness of group
elements and Eq.~\eqref{eq:12} we know
\begin{equation}
\label{eq:57}
  G(\mathcal{O}_{III}) = G_c(\mathcal{O}_{III}) \otimes G_d(\mathcal{O}_{III}) = e^{i
    \sum_{j} a_{j} \sigma_{y}^{(j)}}   \otimes \mathcal{S}^{S}_{1,2,\cdots,n} \otimes
  \mathcal{P}_{1,2,\cdots,n}.
\end{equation}

\begin{figure}[htbp]
  \centering
  \begin{tikzpicture}
    \begin{scope}[every node/.style={draw, circle, fill=gray!20}]
      \node (n1) at (0, 0) {$\boldsymbol{\gamma}_{1}$};
      \node (n2) at (2, 0) {$\boldsymbol{\gamma}_{2}$};
      \node (n3) at (0, 2) {$\boldsymbol{\gamma}_{3}$};
      \node (n4) at (2, 2) {$\boldsymbol{\gamma}_{4}$};
      \node (n5) at (-2, 1) {$\boldsymbol{\gamma}_{5}$};
      \node (n6) at (4, 1) {$\boldsymbol{\gamma}_{6}$};
      \draw (n1) -- (n2) (n3) -- (n4) (n1) -- (n5)  (n3) -- (n5) (n2) -- (n6)
      (n4) -- (n6);
    \end{scope}
  \end{tikzpicture}
  \caption{A graph showing commutation relations in $\mathcal{O}_{IV,v}$ and
    $\mathcal{O}_{IV,h}$.}
  \label{fig:5}
\end{figure}
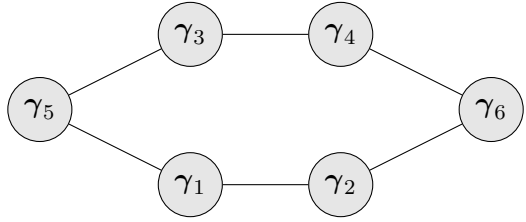

Finally we study an example of the quantum Boltzmann machine with hidden
subsystem, whose Hamiltonian is
\begin{equation}
  H_{IV} = H_{IV,v} + H_{IV,h} + H_{IV,v},
\end{equation}	
where
\begin{align}
  H_{IV,v} & = a_1 \sigma_{x}^{(1)} + a_2 \sigma_{z}^{(1)} + a_3 \sigma_{x}^{(2)} + a_4
             \sigma_{z}^{(2)} +  a_5 \sigma_{z}^{(1)} \otimes \sigma_{z}^{(2)} + a_{6} \sigma_{x}^{(1)} \otimes
             \sigma_{x}^{(2)}, \nonumber\\
  H_{IV,h} & = a_7 \sigma_{x}^{(3)} + a_8 \sigma_{z}^{(3)} + a_9 \sigma_{x}^{(4)} + a_{10}
             \sigma_{z}^{(4)} +  a_{11} \sigma_{z}^{(3)} \otimes \sigma_{z}^{(4)} + a_{12} \sigma_{x}^{(3)} \otimes
             \sigma_{x}^{(4)}, \nonumber\\
  H_{IV,c} & = a_{13} \sigma_{z}^{(1)} \otimes \sigma_{z}^{(3)} +a_{14} \sigma_{z}^{(1)} \otimes
             \sigma_{z}^{(4)} + a_{15} \sigma_{z}^{(2)} \otimes \sigma_{z}^{(3)} + a_{16}
             \sigma_{z}^{(2)} \otimes \sigma_{z}^{(4)} \nonumber\\
           & + a_{17} \sigma_{x}^{(1)} \otimes
             \sigma_{x}^{(3)} + a_{18} \sigma_{x}^{(1)} \otimes
             \sigma_{x}^{(4)} + a_{19} \sigma_{x}^{(2)} \otimes
             \sigma_{x}^{(3)}  + a_{20} \sigma_{x}^{(2)} \otimes
             \sigma_{x}^{(4)}.
\end{align}
The commutation relations for $\mathcal{O}_{IV,v}$ and
$\mathcal{O}_{IV,h}$ are represented by the same graph shown in
Fig.~\ref{fig:5}.

To find out the symmetric groups of $H_{IV,v}$ and $H_{IV,h}$, we first need to
calculate the continuous subgroups and the discrete subgroups respectively. By
solving Eqs.~\eqref{eq:38}\eqref{eq:40}, we find that
$G_{c}(\mathcal{O}_{IV,v})=G_{c}(\mathcal{O}_{IV,h})=I$. By a similar procedure
we obtain the discrete subgroups:
\begin{align}
  G_{d}(\mathcal{O}_{IV,v})
  & = \{ I, \text{H}_{1}\text{H}_{2},
    \text{C}_{12}, \text{H}_{1}\text{H}_{2}
    \times \text{C}_{12},
    (\text{H}_{1}\text{H}_{2} \times
    \text{C}_{12})^2,
    (\text{H}_{1}\text{H}_{2} \times
    \text{C}_{12})^3  \} \times \{ I,
    \text{S}_{12} \} \times \mathcal{P}_{1,2},  \\
  G_{d}(\mathcal{O}_{IV,h}) &  = \{ I, \text{H}_{3}\text{H}_{4}, \text{C}_{34}, \text{H}_{3}\text{H}_{4} \times \text{C}_{34}, (\text{H}_{3}\text{H}_{4} \times \text{C}_{34})^2, (\text{H}_{3}\text{H}_{4} \times \text{C}_{34})^3  \} \times \{ I, \text{S}_{34} \} \times \mathcal{P}_{3,4},
\end{align}
where $\text{H}_{i}$ is the Hadamard gate for qubit-$i$, $\text{C}_{ij}$ is the
controlled-NOT with qubit-$i$ as the control qubit and with qubit-$j$ as the
target qubit. By acting
$G_{d}(\mathcal{O}_{IV,v}) \otimes G_{d}(\mathcal{O}_{IV,h})$ on $H_{IV,c}$, we
obtain the transformations satisfying Eq.~\eqref{eq:47}:
\begin{equation}
  \begin{split}
    G(\mathcal{O}_{IV}) = & \{ I, I_{12} \otimes
    \text{H}_{3}\text{H}_{4}, \text{S}_{12} \otimes
    \text{S}_{34}, \text{S}_{12} \otimes
    (\text{H}_{3}\text{H}_{4} \times \text{S}_{34}),   \text{H}_{1}\text{H}_{2} \otimes I_{34}, \\
    & \text{H}_{1}\text{H}_{2} \otimes \text{H}_{3}\text{H}_{4},
    (\text{H}_{1}\text{H}_{2} \times \text{S}_{12}) \otimes
    \text{S}_{34},
    (\text{H}_{1}\text{H}_{2} \times \text{S}_{12}) \otimes  (\text{H}_{3}\text{H}_{4} \times \text{S}_{34}) \} \\
    & \times \mathcal{P}_{1,2} \otimes \mathcal{P}_{3,4}.
  \end{split}
\end{equation}

\subsection{Numerical verification}
\label{sec:numer-verif}

In Section~\ref{sec:basic-equat-symm}, we give the basic equations for
calculating the symmetry group $G(\mathcal{O})$ of a quantum Boltzmann machine
$\mathcal{O}$. And we also present a general procedure to construct the symmetry
group $G(\mathcal{O})$ in Section~\ref{sec:cont-discr-subgr} with the assumption
that the discrete subgroup $G_{d}(\mathcal{O}_{v})$ is a subgroup of the
Clifford group. Now we numerically calculate the symmetry group
$G(\mathcal{O}_{v})$ with the basic equations, and attempt to show that the
assumption is valid.

The main steps of the numerical algorithm is as follows. First, for a given type
of quantum Boltzmann machine, we can obtain the basic equations
$\{f_{i}(\mathcal{U})=0\}$ from
Eqs.~\eqref{eq:45}\eqref{eq:17}\eqref{eq:46}\eqref{eq:16}. Second, we
numerically solve these basic equations by the Levenberg–Marquardt (LM)
algorithm~\cite{NW2006}. The LM algorithm converts the problem of solving
equations into a nonlinear least squares problem:
\begin{equation}
  \label{eq:21}
  \mathop{\text{min}}_{\mathcal{U}} F(\mathcal{U})   
\end{equation}
with
\begin{equation}
  \label{eq:22}
  F(\mathcal{U}) =   \frac{1}{2} \sum_{i} |f_{i}(\mathcal{U})|^2 .  
\end{equation}
The LM algorithm is a nonlinear optimization between gradient descent and Newton
method. Since $F(\mathcal{U})$ may have many local minima, we often arrive at a
local minimum but not a solution of the equations by the algorithm. To ensure a
solution is arrived at, we need to check the condition that the limit of
$F(\mathcal{U})$ is zero, which is given numerically by
$F(\mathcal{U}^{\star}) \le 10^{-6}$, where $\mathcal{U}^{\star}$ is the
convergent result obtained from the LM algorithm. When our system consists of a
small number $n$ of qubits (here the small number $n$ is related to numbers of
variables and equations, for examples, $n \le 3$ in the cases of $H_{II}$ and
$H_{III}$), we can directly numerically solve the basic equations by the LM
algorithm. When our system consits of more qubits (for examples, $n = 4$ in the
cases of $H_{II}$ and $H_{III}$), we can not solve the basic equations directly.
We find that when
$\forall \gamma^{\prime\prime}, \gamma^{\prime\prime\prime} \in
D(\mathcal{O}_{v}), \exists! \gamma^{\prime\prime} \oplus
\gamma^{\prime\prime\prime} = \gamma_{\star} \in D^{n_{v}} $, one of the basic
equations, Eq.~\eqref{eq:17}, can be simplified to
$\mathcal{U}_{\boldsymbol{\gamma} \boldsymbol{\gamma}^{\prime\prime}}
\mathcal{U}_{\boldsymbol{\gamma} \boldsymbol{\gamma}^{\prime\prime\prime}}=0
\;(\forall \boldsymbol{\gamma}\in D_{a}(\mathcal{O}_{v}), \forall
\gamma^{\prime\prime}, \gamma^{\prime\prime\prime} \in D(\mathcal{O}_{v}) \
\text{and} \ \gamma^{\prime \prime} < \gamma^{\prime \prime \prime} )$, which
implies that
$\mathcal{U}_{\boldsymbol{\gamma} \boldsymbol{\gamma}^{\prime\prime}}=0$ or
$\mathcal{U}_{\boldsymbol{\gamma} \boldsymbol{\gamma}^{\prime\prime\prime}}=0$.
So we first solve these equations and get some possible zero solutions. After
substituting these zero solutions into the rest of the basic equations, we apply
the LM algorithm to solve the simplified basic equations. The process of our
numerical solution of the basic equations is shown in the algorithm~\ref{NSY}.

\begin{algorithm}
  \caption{ Numerical solution of symmetric group }
  \label{NSY}
  \SetNlSkip{0.25em} \SetInd{0.5em}{1em} \LinesNumbered \KwIn{Given the quantum
    Boltzmann machine model: number of qubits $n$, type of Hamiltonian $H$ (eg.
    $H_{II} = \sum_{i=1}^{n} \left(a_{2i-1} \sigma_{x}^{(i)} + a_{2i}
      \sigma_{z}^{(i)}\right) + \sum_{i<j}^{n} a_{i,j} \sigma_{z}^{(i)} \otimes
    \sigma_{z}^{(j)}$), $\gamma$;} \KwOut{Transformation matrices $\mathcal{U}$,
    as show in Eqs.~\eqref{eq:49};} \textbf{Initialize} Random initialization:
  $\mathcal{U}_{\gamma \gamma^{\prime}} \in [-1,1]$;
	
  Obtain the basic equations $\{ f_{i}(\mathcal{U})=0 \}$ from
  Eqs.~\eqref{eq:45}\eqref{eq:17}\eqref{eq:46}\eqref{eq:16};
	
  Solve the basic equations:
	 
  \If{$n \le 3$} {Solve the equations by the LM algorithm;}
	
  \Else { \If {
      $\exists! \gamma^{\prime} \oplus \gamma^{\prime\prime} = \gamma_{\star}
      (\gamma^{\prime}, \gamma^{\prime\prime} \in D(\mathcal{O}_{v}))$} { Obtain
      the equations
      $ \{\mathcal{U}_{\boldsymbol{\gamma} \boldsymbol{\gamma}^{\prime}}
      \mathcal{U}_{\boldsymbol{\gamma} \boldsymbol{\gamma}^{\prime\prime}}=0 \}
      (\forall \boldsymbol{\gamma}\in D_{a}(\mathcal{O}_{v}), \forall
      \gamma^{\prime}, \gamma^{\prime\prime} \in D(\mathcal{O}_{v}),
      \gamma^{\prime } < \gamma^{\prime\prime} )$ from Eq.~\eqref{eq:17};
	 
      \For{all $\gamma^{\prime}, \gamma^{\prime \prime}$} { Find all possible
        zero solutions of
        $\{ \mathcal{U}_{\gamma \gamma^{\prime}}\mathcal{U}_{\gamma
          \gamma^{\prime \prime}} = 0 \}$;}
	 
      \For{all $\gamma$} { Find all the combinations of zero solutions
        $\{ \mathcal{U}_{\gamma \gamma^{\prime}}^{i} | \mathcal{U}_{\gamma
          \gamma^{\prime}}^{i}=0 \}$:
        $C=\{ \{ \mathcal{U}_{\gamma \gamma^{\prime}}^{1} \},
        \{\mathcal{U}_{\gamma \gamma^{\prime}}^{2} \}, \dots \}$; }
	 
      \For{ batches of $i$} {Substitute
        $\{ \mathcal{U}_{\gamma \gamma^{\prime}}^{i} | \mathcal{U}_{\gamma
          \gamma^{\prime}}^{i}=0 \}$ into the basic equations to get the
        simplified basic equations;

        Solve the simplified basic equations with random initial value:
        $\mathcal{U}_{\gamma \gamma^{\prime}} \in [-1,1], \ \mathcal{U}_{\gamma
          \gamma^{\prime}} \notin \{ \mathcal{U}_{\gamma \gamma^{\prime}}^{i} |
        \mathcal{U}_{\gamma \gamma^{\prime}}^{i}=0 \}$; } } }
  
  \textbf{return} Transformation matrices $\mathcal{U}$.
	 
\end{algorithm}

The numerical results on the symmetry groups of the quantum Boltzmann machines
with Hamiltonian $H_{I}$, $H_{II}$ with $n=3,4$, and $H_{III}$ with $n=2,3$ are
shown in Table~\ref{H1},~\ref{H2},~\ref{H3} respectively. Because every term in
our quantum Boltzmann machines is invariant under the operations in their Pauli
groups, the Pauli group is the subgroup of the symmetry groups of our quantum
Boltzmann machines. Here we thus need to consider only the discrete part of the
symmetry groups of our quantum Boltzmann machines beyond the Pauli groups. The
case of Hamiltonian $H_{I},n=2$, as shown in Table~\ref{H1}, we take $500$
different initial values to solve the basic equations of the symmetry group. We
get $\mathcal{U}=I$ for $38$ times, $\mathcal{U}=S_{12}$ for $33$ times, and
local minima (NOT solutions) for $429$ times. Similarly, for the case of
Hamiltonian $H_{II}$ with $n=3$ shown in Table~\ref{H2}, we take $50000$
different initial values to solve the basic equations, $49931$ of which fall to
local minima (NOT solutions). All the symmetry operations we obtained are
consistent with Eqs.~\eqref{eq:32}~\eqref{eq:33}. For the case of Hamiltonian
$H_{III}$ with $n=2$, shown in Table~\ref{H3}, we take $10000$ different initial
values to solve the equations, $8855$ of which fall to local minima.
Furthermore, we get the continuous symmetry $1076$ times and the discrete
symmetries, which are given in Eqs.~\eqref{eq:51}. In summary, we observe that
all the numerical solutions for the symmetry groups have already been obtained
in Section~\ref{sec:cont-discr-subgr}, which shows that our procedure given in
Section~\ref{sec:cont-discr-subgr} is valid for calculating the symmetry group
of a given quantum Boltzmann machine.

\begin{table}
  \caption{ The result of solving the basic equations of the symmetry group for $H_{I},n=2$}
  \setlength{\tabcolsep}{8mm}
  \label{H1}
  \begin{tabular}{| c | c | c || c | c |}
    \hline
    Symmetry & $I$ & $S_{12}$ & local minima & Total InitialValue  \\
    \hline
    Frequency & 38 & 33  & 429  & 500  \\
    \hline
  \end{tabular}
\end{table}

\begin{table}
  \caption{The result of solving the basic equations of the symmetry group for $H_{II},n=3$}
  \setlength{\tabcolsep}{3mm}
  \label{H2}
  \begin{tabular}{| c | c | c | c |c | c |c || c | c |}
    \hline
    Symmetry & $I$ & $S_{132}$ & $S_{213}$ & $S_{231}$ & $S_{312}$ & $S_{321}$ & local minima & Total InitialValue  \\
    \hline
    Frequency & 10 & 10 & 14 & 8 & 12 & 15 & 49931  & 50000  \\
    \hline
  \end{tabular}
\end{table}

\begin{table}
  \caption{The result of solving the basic equations of the symmetry group for $H_{III},n=2$}
  \setlength{\tabcolsep}{0.6mm}
  \label{H3}
  \begin{tabular}{| c | c | c | c |c | c |c | c| c || c|| c | c |}
    \hline
    Symmetry & $I$ & $S_{12}$ & $H_{1}$ & $H_{2}$ & $H_{1}H_{2}$ & $S_{12} \times H_{1}$ & $S_{12} \times H_{2}$ & $S_{12} \times H_{1}H_{2}$ & $G_{c}$ & local minima & Total InitialValue  \\
    \hline
    Frequency & 9 & 6 & 7 & 12 & 11 & 11 & 4 & 9 & 1076 & 8855  & 10000  \\
    \hline
  \end{tabular}
\end{table}

\begin{table}
  \caption{ Running time of our algorithm for different quantum Boltzmann machines}
  \setlength{\tabcolsep}{5mm}
  \label{Time}
  \begin{tabular}{| c | c | c | c |}
    \hline
    & Number of Eqs. & Total InitialValue & Running time (s) \\
    \hline
    $H_{I},n=2$ & 81 & 500  & $4.52 \times 10$ \\
    \hline
    $H_{III},n=2$ & 132 & 10000  & $9.10\times 10^2$ \\
    \hline
    $H_{II},n=3$ & 393 & 50000  & $2.18 \times 10^4$ \\
    \hline
    $H_{III},n=3$ & 1032 & 100000  & $2.08 \times 10^5$ \\
    \hline
    $H_{II},n=4$ & 1298 & 500000 & $1.37 \times 10^6$ \\
    \hline
  \end{tabular}
\end{table}

The running time for the above calculations is shown in Table~\ref{Time}. Our
algorithms are run on 8-core 4.0GHz CPU. With the increase of qubits number, the
number of equations increases rapidly. As the increase of equations, more local
minima of $F(\mathcal{U})$ appear, and we need to take more initial values to
get all the solutions of the basic equations. In the case of $H_{II},n=4$, the
running time is $1.37\times 10^{6}$ seconds in our computer. It implies that it
is almost impossible to calculate the symmetry group for a quantum Boltzmann
machine with more qubits (e.g. a quantum Boltzmann machine with $10$ qubits) in
our computer.

% Among these $500000$ initial values, we randomly select $25000$ combinations
% of possible zero solutions and $20$ initial values for each zero solution to
% solve the new equations.

\section{Discussion and summary}
\label{sec:discussion-summary}

In summary, we introduce the concept of the symmetry of a quantum Boltzmann
machine, and build a general group theory to describe this symmetry. For any
quantum Boltzmann machine composed of multiqubits, we give the basic equations
for its symmetry group. Further more, we classify the symmetry group into the
continuous subgroup and the discrete subgroup, and simplify the basic equations
to the isomorphic graphs with the generators and the same rule to generate all
other elements. To demonstrate how to apply the method, we analyze the
symmetries for four typical quantum Boltzmann machines and obtain their symmetry
groups, which is supported by the numerical calculation based on the basic
equations.

Our work builds a direct connection between group theory and quantum
Boltzmann machine, which answers not only what target states are
equivalent to a quantum Boltzmann machine, but also what solutions are
equivalent to the quantum Boltzmann machine. We expect that such type
of symmetry in a quantum Boltzmann machine can be further extended to
study the symmetries in other machine learning models, which can be
used to obtain their global symmetric features in simulating physical
data.

\begin{acknowledgments}
  This work is supported by NSF of China (Grant Nos. 11475254 and
  11775300), NKBRSF of China (Grant No. 2014CB921202), the National
  Key Research and Development Program of China (2016YFA0300603).
\end{acknowledgments}

\bibliography{symqbm}

\end{document}